\documentclass[12pt]{article}

\usepackage{xcolor}

\usepackage{scicite}
\usepackage{amsmath} 
\usepackage[utf8]{inputenc}
\usepackage{graphicx}
\usepackage{float}

\usepackage{times}

\newenvironment{sciabstract}{%
\begin{quote} \bf}
{\end{quote}}

\newcounter{lastnote}

\title{
High-Accuracy Dispersion Surfaces Measurement Unlocks Photonic Crystal Slabs Topology}

\author
{Karen Caicedo, $^{1}$ Fabrizio Sgrignuoli,$^{1}$ Adam Schwartzberg,$^{2}$, Scott Duehy,$^{2}$ \\
Silvia Romano, $^{1}$ Gianluigi Zito, $^{1}$ Ivo Rendina, $^{1}$  Vito Mocella $^{1\ast}$ \\
\\
\normalsize{$^{1}$Institute of Applied Sciences and Intelligent Systems, National Research Council, Naples, Italy}\\
\normalsize{$^{2}$Molecular Foundry, Lawrence Berkeley National Laboratory, Berkeley, CA, USA}\\
\\
\normalsize{$^\ast$E-mail: vito.mocella@na.isasi.cnr.it}
}

\date{}

\begin{document} 


\baselineskip24pt


\maketitle


\begin{sciabstract}

 Characterization of dispersion surfaces (DS) in photonic crystals (PhCs) can predict striking topological features, such as bound states in the continuum (BICs). Precise measurement of dispersion, particularly near the $\mathbf{\Gamma}$-point, is crucial since even subtle geometric deviations significantly impact the distribution of topological charges and associated polarization singularities. Here, we propose a novel technique to measure DS of PhC slabs with high angular accuracy of up to $\mathbf{3 \times10^{-4}}$ radians, and spectral accuracy of 0.2 nm within an expanded region of the Brillouin zone. Remarkably, our technique is diffraction- and aberration-free as it does not rely on optical imaging but rather on spectral mapping. The analysis of DS presented in this work demonstrates the robustness of our technique, and enables us to distinguish anisotropic BICs along a specific symmetry axis from isotropic states around the $\mathbf{\Gamma}$-point. Moreover, the experimental DS obtained supply the information necessary to study and identify super-BICs.

\end{sciabstract}

\section*{Introduction}
Topology has emerged as a transformative concept in solid-state physics as well as photonics, offering profound insights into the design and manipulation of light-matter interactions beyond traditional electromagnetic constraints \cite{lu2014topological}.
By leveraging topological invariants, photonic systems can host protected edge states and unidirectional propagation \cite{Raghu:2008do, 10.1126/sciadv.aaw4137}, enabling unprecedented control over electromagnetic waves.
This paradigm extends naturally to photonic crystals (PhCs),  where periodic dielectric structures exploit topology to tailor photonic bandgaps and guide light with topological protection \cite{Wu:2015ex}.
Particularly intriguing are PhCs supporting bound states in the continuum (BICs)\cite{hsu2013}, which represent a unique class of topological states \cite{Zhen:2014eo} characterized by their ability to localize electromagnetic energy within a nominally radiative spectrum.

Conventionally, both theoretical and experimental studies of PhCs restricted their focus to high-symmetry directions of the Brillouin zone (BZ). However, off-symmetry regions have been overlooked, despite the fact that they exhibit surprisingly rich physics. Among these are self-collimation beams \cite{KosakaAppliedPhysicsLetters1999, Prather2007, Mocella:2009dr}, zero-group-velocity modes \cite{Notomi:2000uq, ToshihikoNatPhot2008}, and exotic topological singularities \cite{10.1038/nphys4072, Zhen:2014eo}, all of which risk going unnoticed when observations are confined to a few discrete wave vectors. So-called anisotropic BICs have been demonstrated in PhC slabs, as well as accidental BICs protected by in-plane symmetry \cite{li2019symmetry, LeePRL2012, SchiattarellaNature2024, ZhangPRL2025}. Although high-symmetry measurements can locate some of these states, off-symmetry explorations are often needed to capture their full momentum-space dispersion.

A more complete perspective emerges when we consider the entire BZ, notably through the mapping of dispersion surfaces (DS). DS can be studied through equifrequency surfaces (EFS). EFS — DS slices in reciprocal space at fixed frequencies — are crucial for understanding the topological properties of far-field polarization states\cite{Zhen:2015bl}. For instance, experimental measurement of EFS  confirmed  the momentum-space polarization vortices in plasmonic crystals, linked to vanishingly small coupling with radiation channels  \cite{zhang2018observation}. Moreover, the same comprehensive approach to DS mapping is central to uncovering other topological photonic phenomena such as optical Weyl points — the three-dimensional analogs of Dirac points — which exhibit nontrivial Chern numbers and support “Fermi arc”–like surface states \cite{Zhou:2018dy}. Detailed mapping of the entire BZ and its EFS enables direct detection of these topological singularities, as well as the arcs that connect them \cite{10.1038/nphys4072, li2018weyl, PhysRevA.93.061801}.
  
Near the $\mathbf{\Gamma}$-point, precise measurements of DS is critical: even minimal geometric deviations can modify topological charge distributions and their associated polarization singularities \cite{Zhou:2018dy}. High accuracy ensures reliable engineering of topologically robust polarization phenomena, enabling devices that harness BICs’ unique properties while tracking polarization texture dynamics near regions of charge creation/annihilation.
This is particularly relevant in the case of so-called super-BICs, where the superposition of various BICs converges in the reciprocal space and produces a bound state whose Q-factor increases much faster than in the case of conventional BICs \cite{jin2019,zhang2025,hwang2021ultralow}, exhibiting flat EFS around the super-BIC \cite{le2024}.
  
Despite the demonstrated importance of DS when understanding topological effects, traditional Fourier optics have been the only methodology available in literature behind EFS imaging, wherein the back-focal plane of an objective lens is imaged onto a detector or spectrometer \cite{lan2023visualization, nguyen2023direct, regan2016direct, zhang2018observation}. Recent works have used in situ photoluminescence or cross-polarization reflection to obtain EFS \cite{lan2023visualization, nguyen2023direct, regan2016direct}. Further refinements have shown that weak scattering from intrinsic disorder can map out EFS \cite{regan2016direct}.  Nevertheless, these existing far-field methods often face practical constraints that limit angular and spectral resolution. For example, we have estimated that typical Fourier optics setups achieve an angular resolution of $\sim 10^{-2}$ radians \cite{regan2016direct, zhang2018observation, Zhou:2018dy}, and band-pass filters commonly used for equifrequency snapshots may have bandwidths of 5-10 nm. These methods typically rely on scanning one angle while filtering over wavelength, an approach that is inherently much less accurate than direct spectral measurements with a spectrometer. Not to mention the fact that Fourier optics setups are constrained by the diffraction limit as well as optical aberrations, resulting mainly in qualitative conclusions.
  
In this work, we present a high-accuracy technique to measure and analyze the EFS of PhC slabs, attaining an angular resolution of the order of $10^{-5}$ radians and a spectral resolution of 0.2 nm over a 430 nm wide spectrum, surpassing common implementations by orders of magnitude, with the added advantage of being fully automated. Notably, our approach is diffraction- and aberration-free as it does not rely on optical imaging, but rather on spectral mapping in the reciprocal space.
By drawing an analogy between photonic dispersion surfaces and Fermi surfaces in condensed-matter physics, the method we propose can be regarded as the photonic equivalent of angle-resolved photoemission spectroscopy (ARPES) \cite{ARPES2022, damascelli2003}.
Our technique, using full 2D angle-resolved transmission spectroscopy, provide direct high-resolution mapping of the photonic DS, in contrast to Fermi surface measurements using ARPES, which are often limited by material purity or thermal broadening \cite{ARPES2022}.
The high-accuracy measurements we performed successfully revealed the underlying topological features of the PhC sample with unprecedented clarity. In particular, we analyze flat EFS, signature expected to observe in a super-BIC.

\section*{Results}
\paragraph*{High-accuracy measurement of dispersion surfaces}
\hfill \break
The technique we developed for EFS quantification and visualization consists of performing bi-angle-resolved transmission spectroscopy on PhC samples, eliminating the restrictions of diffraction limit and optical aberrations inherently present in optical imaging methodologies, such as Fourier optics. For this purpose, we have designed a simple and compact, yet powerful and clever, optical setup (Fig. 1.A) composed by a supercontinuum laser source (SCS) that traverses a Glan-Thompson (GT) prism to select the polarization desired for characterization, either s or p (or any other polarization). After, the laser impinges the sample, which is placed in a stage that rotates in both polar ($\theta$) and azimuthal ($\phi$) directions (Fig. 1.B). Finally, the transmission resulting from the coupling between the laser and the PhC resonances is measured using a high-resolution spectrometer as we move in the ($\theta$, $\phi$) space. The core principle behind this technique relies on the fact that polar rotation covers different incidence angles, allowing us to explore optical features as we move further away from the $\Gamma$-point in a given direction. Meanwhile, azimuthal rotation enables us to explore not only the high-symmetry axes, such as $\Gamma$X and $\Gamma$M, but the full region around the $\Gamma$-point. These rotations cover a circular region within the BZ, as visually explained in Fig. 1.C. After the acquisition process, the data is filtered to subtract the background signal. Then, the signal originally recorded in the real ($\theta$, $\phi$) space is transformed and mapped onto the reciprocal ($k_x$, $k_y$) space for each wavelength measured by the spectrometer, where EFS are formed. The beam diameter has a size comparable to the sample dimensions (1 mm $\times$ 1 mm),  which determines the maximum attainable momentum-space resolution ($6.28\times 10^{-3}$ $\mu \text{m}^{-1}$). The angular accuracy experimentally achieved with our method is $3 \times 10^{-4}$ radians, while the spectral accuracy is 0.2 nm, which combined corresponds to $5.1 \times 10^{-3}$ $\mu \text{m}^{-1}$ in the momentum-space, comparable to the maximum resolution determined by sample/beam size. The spectral resolution can easily be improved by changing the grating inside the spectrometer, while reducing the spectrum bandwidth. It should be emphasized that, experimentally, the resolution attained is limited by the windowing of the sample as the maximum real‐space dimension effectively sets the finest “grid” or resolution in the reciprocal space. However, if we would dispose of bigger samples (and adjusting the beam size to them), we could reach an angular resolution of the order of $10^{-7}$ radians.

\begin{figure}[H]
    \centering
    \includegraphics[width=0.5\textwidth]{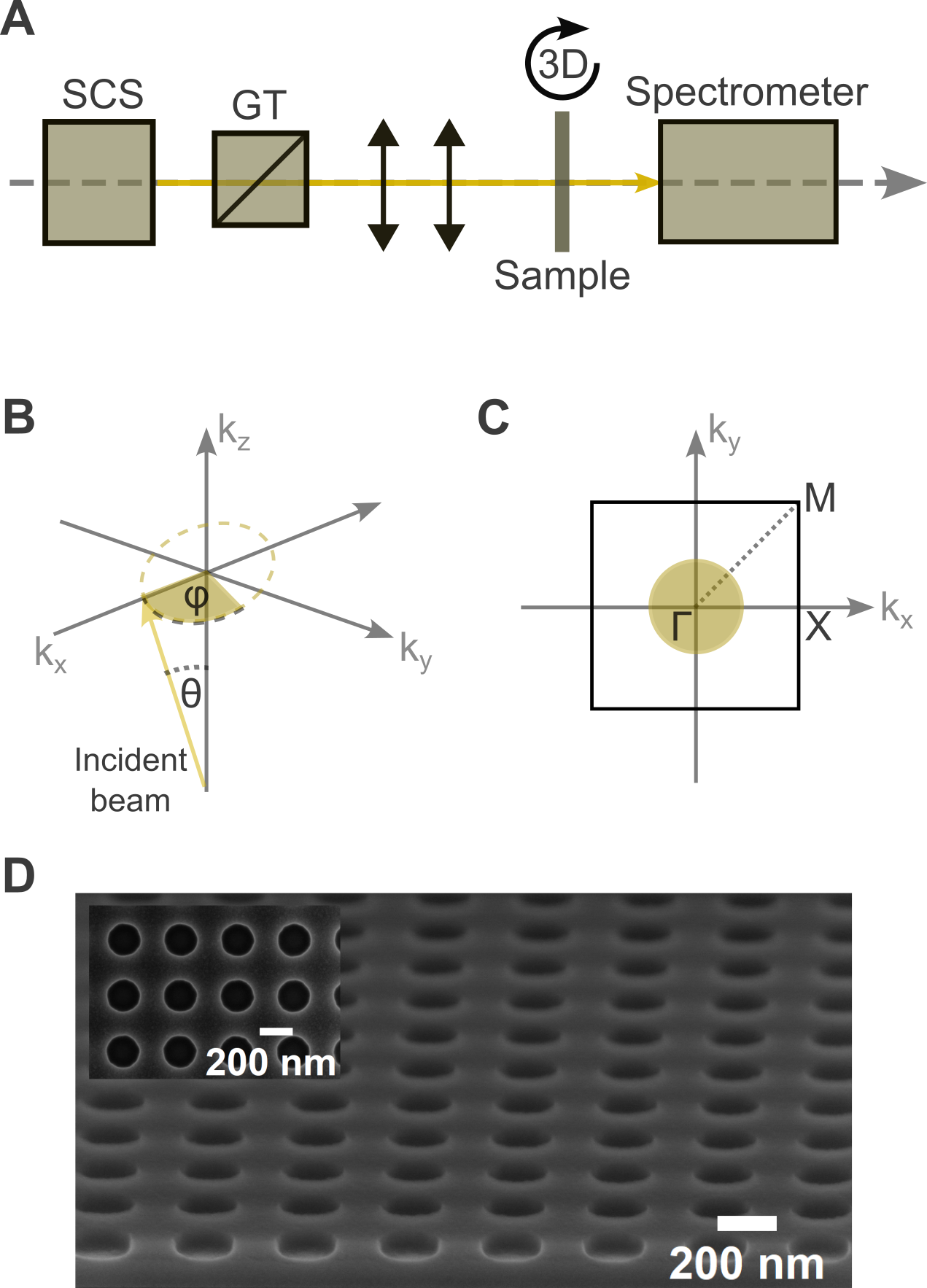}
    \caption{\textbf{Optical setup, reciprocal space, and PhC sample structure.} (A) Schematics of the experimental setup employed for achieving EFS visualization. It consists of a supercontinuum laser source (SCS) that passes through a Glan-Thompson (GT) prism, reaches the sample placed in a holder that allows bi-angular rotations, and the signal is finally measured by a high-resolution spectrometer. (B) Polar ($\theta$) and azimuthal ($\phi$) angles in relation to the reciprocal space. (C) First BZ of a square periodic lattice. The directions $\Gamma$X and $\Gamma$M are known as high-symmetry axes. The circular area corresponds to the space covered during the measurements for the obtention of EFS. (D) Scanning electron microscopy images of the PhC sample.}
    \label{fig:system}
\end{figure}

This method has been demonstrated by using a PhC slab square lattice with periodic cylindrical air holes ( Fig. \ref{fig:system}.D). See Methods for more details about the fabrication methodology. The lattice parameters were adjusted to achieve a super-BIC at around 570 nm in the visible range of wavelengths. The periodicity of the lattice was set to $a = 370$ nm, the radius of the lattice holes to $r = 0.25a$, while the thickness of the PhC membrane was fixed to $h = 0.22a$. These values were determined following the method outlined in reference \cite{Mocella2015}.

Our experimental findings are presented in Fig. 2. EFS for an s-polarized light source are found in Fig. 2.A, where we observe the evolution of the EFS as we move along different wavelengths (517.6, 527.7, 532.0, 539.8, 546.1, 549.9, 554.3, and 589.1 nm). The EFS correspond to the white structures in the images. The complex shapes and the interplay between bands evidence the outstanding performance of our method. Remarkably, the resolution achieved in this experiment is $5.1 \times 10^{-3}$ $\mu \text{m}^{-1}$, which corresponds to the minimum possible resolution limited by the sample dimensions. As the resolution is tuned by choosing the angular step, we have performed a finer experiment, where the spacing realized in the momentum space is $5.1 \times 10^{-4}$ $\mu \text{m}^{-1}$. However, as this spacing is smaller than the experimental resolution permitted, any features narrower than roughly $6 \times 10^{-3}$ $\mu \text{m}^{-1}$ begin to blend together and not be resolved. An example of the EFS obtained in this set of experiments is indicated in Fig. 2.A as a close-up of the region around $\Gamma$ at 532.0 nm. Fig. 2.B displays the EFS for a p-polarized light source as we explore the same wavelengths with equal resolution as for s-polarization.

\begin{figure}[H]
    \centering
    \includegraphics[width=0.9\textwidth]{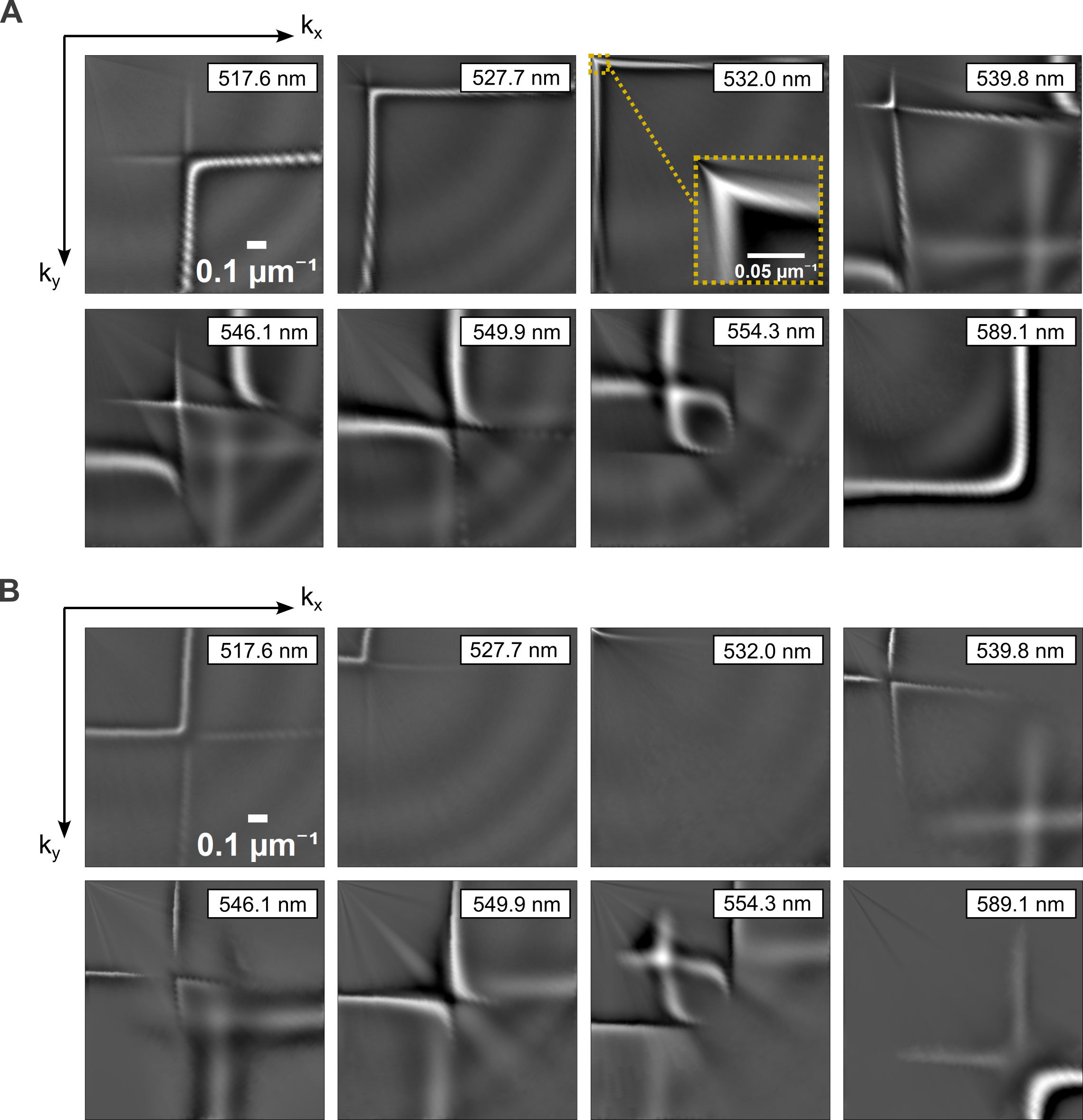}
    \caption{\textbf{High-resolution equifrequency surfaces visualization.} (A) EFS of a PhC sample impinged by an s-polarized broadband laser source. EFS are imaged in the first quadrant of the reciprocal space, where $\Gamma$ sits at the origin of the coordinates system. Different wavelengths were explored: 517.6, 527.7, 532.0, 539.8, 546.1, 549.9, 554.3, and 589.1 nm. The scale bar in the images corresponds to 0.1 $\mu \text{m}^{-1}$, and the resolution achieved is $5.1 \times 10^{-3}$ $\mu \text{m}^{-1}$. At 532.0 nm, the EFS of a finer experiment was included as a zoomed-in view of the closer region around $\Gamma$. Here, the scale bar corresponds to 0.05 $\mu \text{m}^{-1}$, and the spacing achieved is $5.1 \times 10^{-4}$ $\mu \text{m}^{-1}$, beyond the resolution experimentally permitted. (B) EFS of a PhC sample impinged by a p-polarized broadband laser source with equal resolution as (A).}
    \label{fig:mesh1}
\end{figure}

Stacking the EFS leads to a 3D visualization of the complete photonic behaviour of the PhC throughout a broad range of frequencies, as shown in Fig. 3.A for both s- and p-polarization. The orthogonal cuts observed correspond to the wavelengths 532.0 and 589.1 nm. The 3D DS unveils the intricate band structure, making it possible to explore in detail unique photonic effects, such as isotropic and anisotropic BICs, as well as self-collimation effects.

\begin{figure}[H]
    \centering
    \includegraphics[width=1\textwidth]{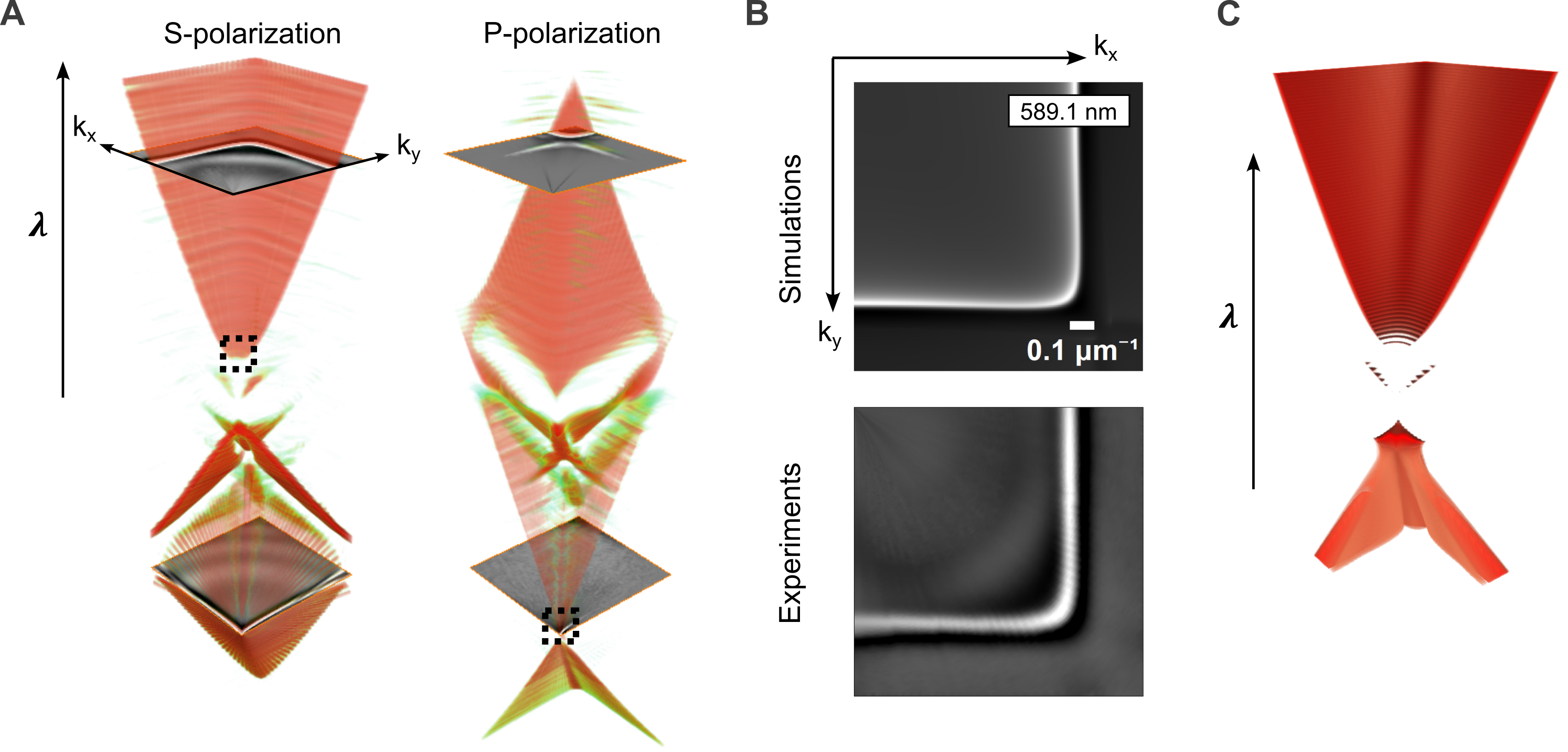}
    \caption{\textbf{3D dispersion surfaces and numerical verification.} (A) 3D landscapes of photonic bands in the reciprocal space throughout a broad range of wavelengths ($\lambda$) going from 500.6 to 597.7 nm. s- and p-polarized light source were used, and the 3D bands are shown side by side. Orthogonal cuts are included at 532.0 (lower) and 589.1 nm (upper).  The dashed squares indicate regions where BICs are found. BICs are discussed in the following section of this article. A zoomed-in-view of these regions are shown in Fig. 4.A and B. (B) Simulated and experimental EFS at 589.1 nm for s-polarization. (C) Simulated 3D DS for s-polarization.}
    \label{fig:3DandSimul}
\end{figure}

Our results have been compared with numerical simulations (see Fig. \ref{fig:3DandSimul}.B) of EFS performed using a custom-made, fully parallelized Python implementation of the rigorous coupled-wave analysis (RCWA) method to solve Maxwell's equations for periodic structures by representing the electromagnetic fields and material permittivity as Fourier harmonics, optimized specifically for high-performance computing (HPC) environments \cite{Moharam:86, Xu_SPIE:2023}. See Methods for more details about the computational methodology. Fig. \ref{fig:3DandSimul}.B shows the EFS at 589.1 nm for s-polarization, revealing perfect agreement between simulations and experiments. Fig. \ref{fig:3DandSimul}.C displays the simulated 3D DS for s-polarization in a region smaller than that obtained experimentally and demonstrates the agreement between simulations and experiments in a wide range of wavelengths.

\paragraph*{Isotropic and anisotropic BICs}
\hfill \break
As it is well known, BICs are topological phenomena. A typical scenario associated with a symmetry-protected BIC at the $\Gamma$-point is that its polarization remains undefined, as it emits no radiation into the far field. Consequently, the polarization vector field around such BIC exhibits a vortex structure characterized by a unitary topological charge, which describes how the polarization vector winds around the BIC \cite{Zhen:2014eo}. Around the $\Gamma$-point, the vortex winding, that defines topological charges, is connected to point-group symmetries of the system — such as mirror reflections and rotations —. In such cases, one would expect a fully symmetric shape of the three-dimensional DS around the $\Gamma$-point at the BIC frequency and DS have the typical shape of Dirac cone\cite{ardizzone2022}, exhibiting a perfectly circular EFS around the BIC, which corresponds to the horizontal cut of the Dirac cone.
However, we demonstrate below that the level of isotropy of the BIC is not necessary correlated with the isotropic degree of the PhC slab, because we have found both isotropic and anisotropic BICs in the same squared PhC lattice. We underline that a vertically asymmetric PhC slab is not isotropic, because a square lattice does not exhibit a high level of rotational symmetry around the $\Gamma$-point, and the vertical asymmetry induces mode hybridization.

Until now, experimental studies on BICs have been essentially limited to band diagrams along symmetric directions. However, this approach overlooks the features of BICs in the rest of the BZ. For instance, the isotropic degree of BICs cannot be fully assessed by only looking at high-symmetry axes. Instead, a study around the full BZ is required.
Notably, our experimental method addresses this gap, allowing, for the first time, a comprehensive 3D visualization of an isotropic BIC and a anisotropic BIC with unprecedented resolution. Furthermore, we are able to study in detail the DS in close proximity around the BIC. In the following, we detail our findings on the isotropic and anisotropic behaviors of BICs, observed in the same sample around the $\Gamma$-point under different polarization illumination.

\begin{figure}[H]
    \centering
    \includegraphics[width=1\textwidth]{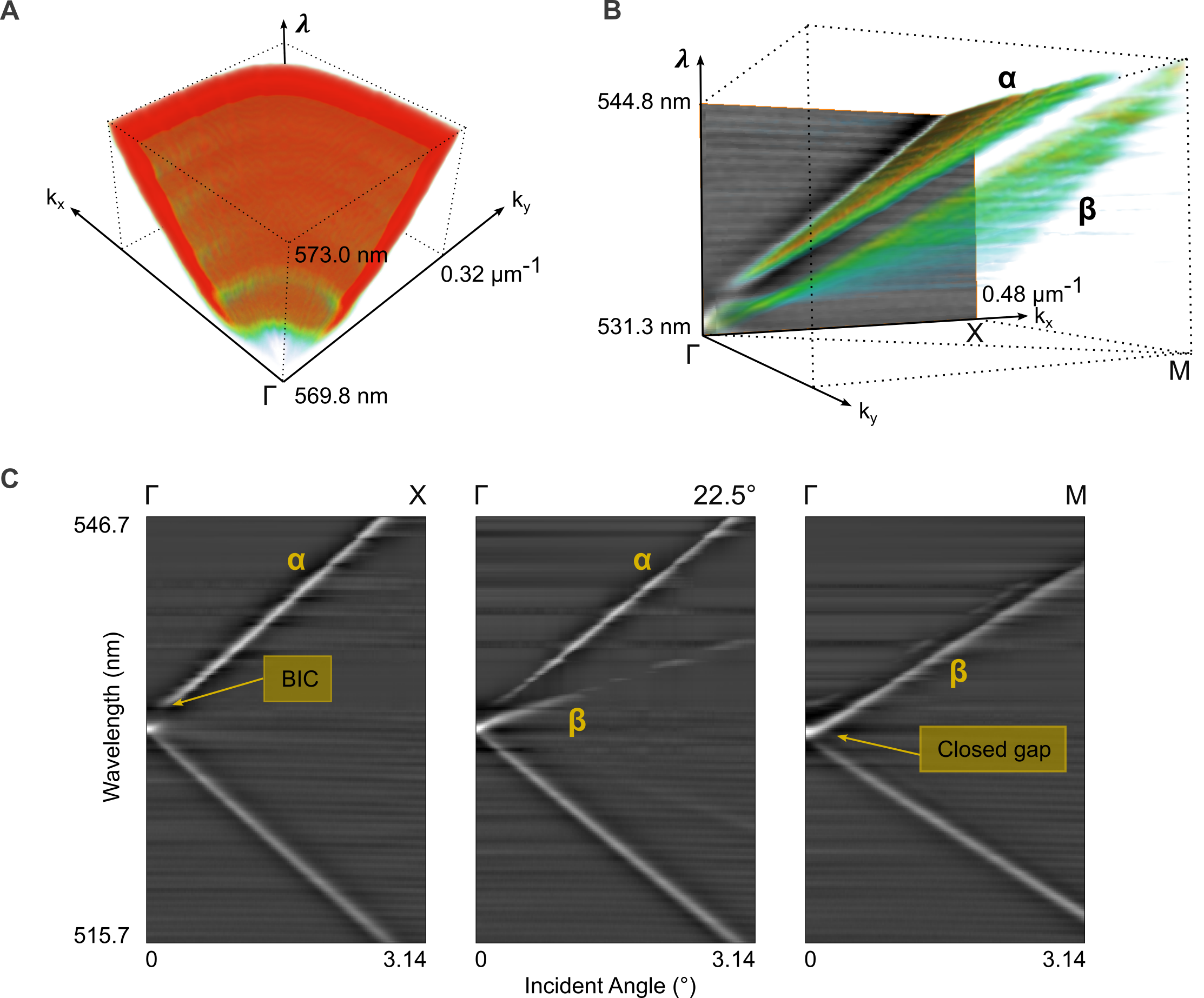}
    \caption{\textbf{Isotropic and anisotropic BICs high-resolution visualization.} (A) 3D dispersion surface vanishes as it approaches the $\Gamma$-point at around 569.8 nm revealing an isotropic BIC under s-polarized laser source. This region corresponds to the one within the dashed square in Fig. 3.A for s-polarization. (B) 3D DS shows a BIC in the $\Gamma$X axis at around 532 nm. The BIC is part of $\alpha$ DS. However, as we move along the azimuthal direction, there is an interplay between surface $\alpha$ and $\beta$, until $\alpha$ vanishes and only surface $\beta$ appears along $\Gamma$M. Remarkably, there is no BIC associated to this surface. This region corresponds to the one within the dashed square in Fig. 3.A for p-polarization. (C) Band diagrams at $\Gamma$X, $\Gamma$-22.5°, and $\Gamma$M, demonstrating the evolution of the anisotropic BIC in (B).}
    \label{fig:BIC_isotropic_anisotropic}
\end{figure}

The 3D DS, obtained with s-polarized input light, revealed an isotropic BIC (Fig. \ref{fig:BIC_isotropic_anisotropic}.A) at around 569.8 nm as the DS vanishes approaching $\Gamma$ from every azimuthal direction, not only following high-symmetry axes. This is a signature of a true symmetry protected BIC. 
Notably, the isotropic BIC also exhibits an isotropic negative effective refractive index — determined by the gradient of the 3D DS — and an isotropic negative effective mass, arising from their convexity \cite{Kittel:371097,Wurdack2023}.

Likewise, we have identified an anisotropic BIC near 532.0 nm for p-polarized input light (see Fig. \ref{fig:BIC_isotropic_anisotropic}.B and C). An interplay between two DS, $\alpha$ and $\beta$, takes place as we move from $\Gamma$X to $\Gamma$M (Fig. \ref{fig:BIC_isotropic_anisotropic}.B). Only the 3D DS reveals such interaction. Along $\Gamma$X, we observe the presence of the BIC associated to the $\alpha$-surface. However, as we move in the azimuthal direction, this band disappears and a band associated with $\beta$-surface appears and reveals no BIC along $\Gamma$M. This progression is shown in Fig. 4.C with three band diagrams along $\Gamma$X, $\Gamma$-22.5°, and $\Gamma$M. Once more, the isotropy and anisotropic conduct of the bands can only be understood in detail through the study of 3D DS in Fig. \ref{fig:BIC_isotropic_anisotropic}.A and B. In this case, the BIC is not symmetry-protected, and its topology is not associated with a vortex at the $\Gamma$-point, which typically corresponds to an integer topological charge \cite{Zhen:2014eo}. Instead, the topology is more complex and involves C-point singularities. Using an interferometric setup at a wavelength of 532.0 nm, we previously investigated the corresponding half-integer topological charge in a sample with the similar parameters as the sample studied in this article \cite{DeTommasi:2023}.


\paragraph*{Super-BIC analysis}
\hfill \break
BICs can be broadly categorized into two types: symmetry-protected and accidental\cite{Hsu:2016dv}. Symmetry-protected BICs typically occur at high-symmetry points in the BZ and arise due to selection rules dictated by the structure’s point-group or mirror operations, which forbid radiative coupling.
Accidental BICs, on the other hand, appear due to parameter tuning and interference effects at points away from high-symmetry locations\cite{Hsu:2016dv,hwang2021ultralow}. When two or more BICs merge, a novel regime known as a super-BIC emerges. Super-BICs display extraordinarily high Q-factors over a broad range of wavevectors, thus offering robust protection against radiation losses and practical advantages in device fabrication and operation \cite{zhang2025,hwang2021ultralow}.


According to ref. \cite{le2024}, a super-BIC is characterized by a significantly high Q-factor that remains unchanged over an extended region of the BZ. This translates into a strong confinement and minimal radiative losses as well as robustness against angular dispersion. High Q-factor is identified by an ultraflat shape of the dispersion curve as it approaches the super-BIC (meaning that it has zero curvature and infinite effective mass) \cite{le2024}.

Before assessing whether BICs are super-BICs or not, we must discuss the asymptotic behaviour of the Q-factor in super-BICs. Following the perturbation theory approach of ref. \cite{zhang2025}, let's consider a BIC happens at frequency $\omega_0$ in the $\Gamma$-point. We can explore the region around the BIC in the reciprocal space by defining $\boldsymbol{\delta} = \delta (\cos \theta, \sin \theta) / a$, where $\delta$ is small enough to perform an expansion in frequency around the BIC:
\begin{equation}\label{eq:expasion}
    \omega = \omega_0 + \delta^2 \omega_2 (\theta) + \delta^4 \omega_4 (\theta) + \ldots
\end{equation}
Due to $C_4$ symmetry, only even terms in the expasion of Eq. \ref{eq:expasion} are non-zero, and Q-factor, defined as $\text{Q} = Re\{ \omega \} / 2 Im \{ \omega \}$,
typically scales to $\sim$ 1/$\delta^2$ because $\omega_0$ is purely real by definition of BIC. Thus, according to perturbation theory, if the Q-factor of resonant states near the BIC scales as Q $\sim$ 1/$\delta^4$ or higher, the BIC is a super-BIC. The shape of the dispersion curve is determined by the lowest order of expansion in Eq. \ref{eq:expasion}. Typically, the EFS around $\omega_0$ are circular because it is dominated by $\delta^2$. Therefore, the criterion used to identify super-BICs in ref. \cite{le2024} is equivalent to study the real part of Eq. \ref{eq:expasion}, where there is a suppression of $Re \{ \omega_2 \}$, making $\delta^4$ the predominant term. We realized that this translates into EFS with ultraflat shape.

\begin{figure}[H]
    \centering
    \includegraphics[width=1\textwidth]{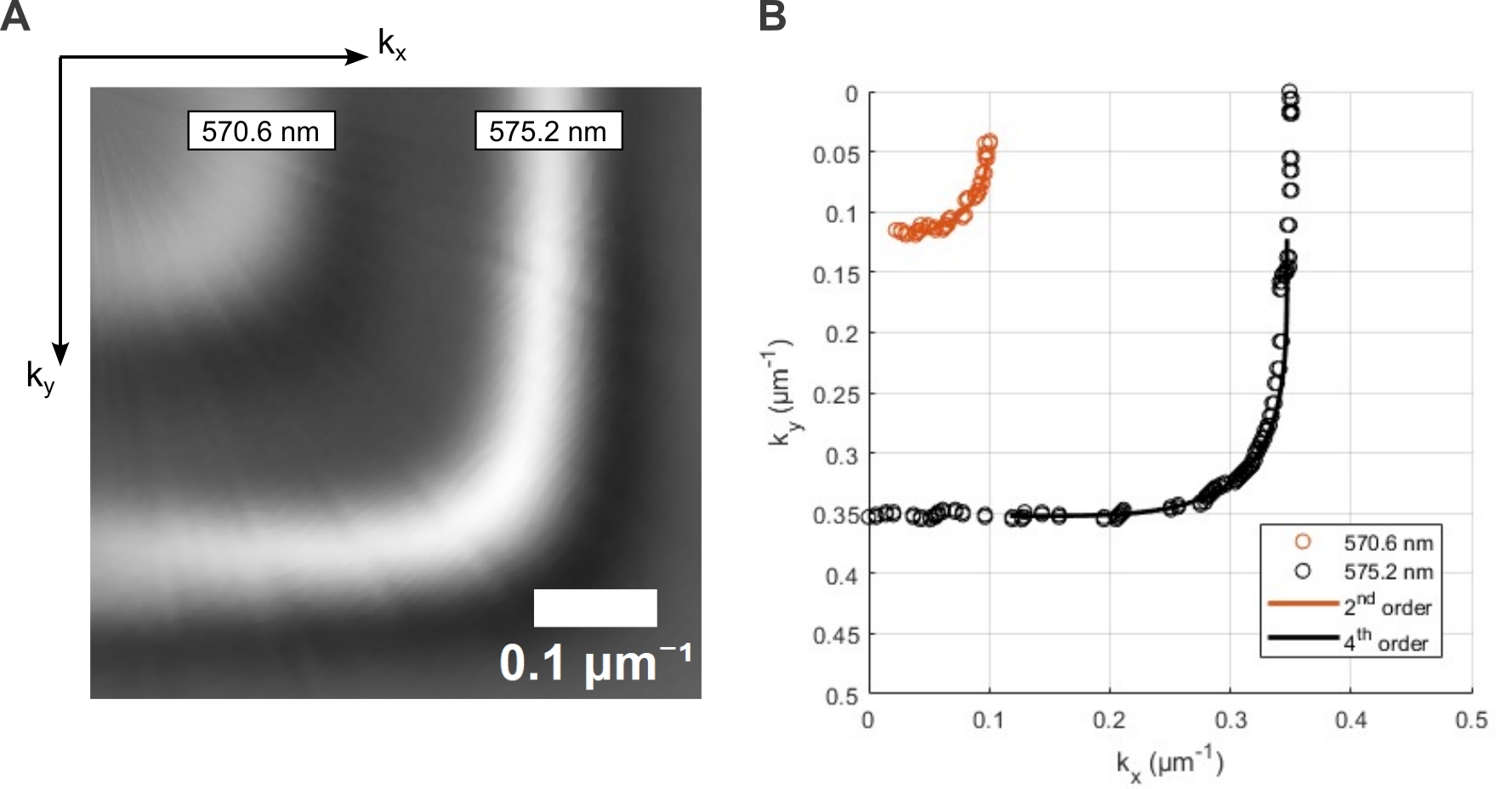}
    \caption{\textbf{Analysis of super-BIC behavior.} (A) EFS at 570.6 and 575.2 nm at approximately 0.1 and 0.35 $\mu \text{m}^{-1}$, respectively, from the isotropic BIC. (B) Fitting (solid lines) of EFS (circular markers) on (A) showing a circular shape of the closer EFS at 570.6 nm, and a flatter shape of the EFS at 575.2 nm. The EFS shapes translate into a quadratic dependency of $(k_x,k_y)$ at 570.6 nm, and a fourth-order dependency of $(k_x,k_y)$ at 575.2 nm.}
    \label{fig:superBIC_final}
\end{figure}

Notably, the measured EFS in the isotropic BIC of Fig. \ref{fig:BIC_isotropic_anisotropic}.A are squared in shape, in contrast to circular EFS of the classic Dirac cone, indicating a suppression of $Re \{ \omega_2 \}$. Due to the remarkable accuracy of our metholody, we are able to distinguish the shape of EFS very close to the $\Gamma$-point. For instance, in Fig. \ref{fig:superBIC_final}.A, we find an EFS at around 0.1 $\mu \text{m}^{-1}$ (corresponding to 570.6 nm), and at around 0.35 $\mu \text{m}^{-1}$ (corresponding to 575.2 nm) from the BIC. However, we have found that the suppression of $Re \{ \omega_2 \}$ is not perfect in the real case since the EFS closer to $\Gamma$ present circular shapes, indicating that in the vicinity of $\Gamma$, 0.1 $\mu \text{m}^{-1}$ and below, the second order becomes predominant in Eq. \ref{eq:expasion} (see Fig. \ref{fig:superBIC_final}.B). Meanwhile, as we move further away, the EFS shape becomes flat (see Fig. \ref{fig:superBIC_final}.B) This denotes that, in our case, there is an almost perfect merging of BICs around the $\Gamma$-point, but a complete merging does not occur. A super-BIC is characterized by a merging of multiple BICs, and our results demonstrate the capability of the proposed methodology to study the merging process of BICs in close detail. In the present case, our methodology revealed that the isotropic BIC is a quasi-super-BICs, due to the almost complete BICs merging.

\begin{figure}[H]
    \centering
    \includegraphics[width=1\textwidth]{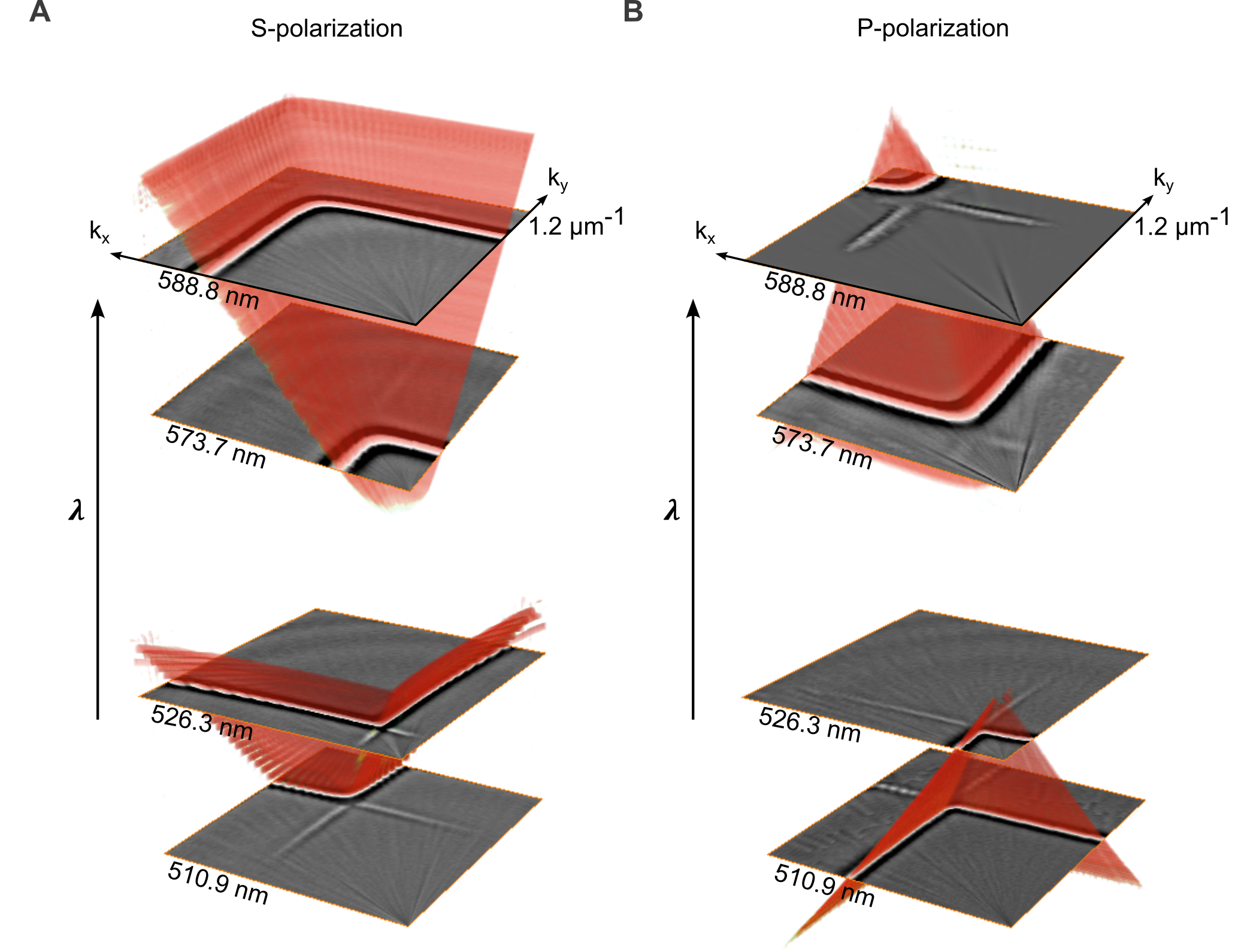}
    \caption{\textbf{Broadband self-collimation 3D visualization.} (A) Two different regions of negative refractive index self-collimation in the 3D DS under s-polarized input source. The orthogonal cuts are included at 588.8, 573.7, 526.3, and 510.9 nm. (B) Same wavelength regions showing positive refractive index self-collimation in the 3D DS under p-polarized input source. Orthogonal cuts occur at the same wavelengths as in (A).}
    \label{fig:self_collimation}
\end{figure}

\paragraph*{Negative refraction self-collimating effect}
\hfill \break
Self-collimation, together with negative refraction, represents one of the earliest phenomena identified in PhCs where EFS analysis plays a key role \cite{KosakaAppliedPhysicsLetters1999, Prather2007, Mocella:2009dr, SchiattarellaNature2024}. In many of these initial studies, the typical focus was theoretical EFS calculations, often in two-dimensional geometries. Experimentally, attention centered primarily on observing the self-collimation or negative refraction effects themselves, rather than directly measuring the underlying EFS. Our method has enabled us to overcome this issue and observe these phenomena through EFS. Remarkably, in Fig. \ref{fig:self_collimation} we observe broadband regions of self-collimation both for negative and positive effective refractive index under s- (Fig. \ref{fig:self_collimation}.A) and p-polarization (Fig. \ref{fig:self_collimation}.B), respectively. Fig. \ref{fig:self_collimation} shows flat-band evolution in the wavelength ranges from 499.0 to 531.5 nm, and from 569.4 to 596.6 nm. EFS are indicated at 510.9, 526.3, 573.7, and 588.8 nm, evidencing self-collimation throughout these regions. To our knowledge, this is the first time that has been experimentally demonstrated the EFS corresponding to negative refractive index self-collimation. In the corner of the EFS, along the $\Gamma$M direction, the refracted beam experiences a superprism effect, characterized by a $90^{\circ}$ deviation in correspondence of a small change in the incident angle \cite{Glockler:2007da, Merzlikin:2006hf, Kosaka:1998vr}, similar to what we have demonstrated in ref. \cite{SchiattarellaNature2024}.

Negative refractive self-collimation can be interesting to explore as the ability to access both self-collimation and negative refraction at one operating frequency can significantly expand design flexibility for integrated optical components through compact, multifunctional photonic devices that can route, focus, enhance local field, or reshape beams in ways not feasible with conventional optical materials.

\section*{Discussion}
The high-accuracy method presented here for measuring EFS in PhCs marks a substantial advancement beyond traditional Fourier optics approaches. By simultaneously acquiring spectral information across both polar and azimuthal angles, we have demonstrated the capability to reconstruct full 3D DS maps with unprecedented resolution.

Our approach has significant implications for understanding topological phenomena in photonic systems, most notably BICs.
Through our method, we have clearly distinguished isotropic, symmetry-protected BICs from anisotropic, accidental BICs in $\Gamma$ by directly observing their full angular dependence.
The unprecedented angular and spectral accuracy obtained has facilitated the first-ever (to our knowledge) experimental observation of 3D DS associated to isotropic BICs around the $\Gamma$-point, enabling a precise examination of their underlying topological nature. Such clarity paves the way for harnessing these states in practical applications like lasers with ultra-high quality factors or robust optical resonators immune to radiative losses.
Our exceptional resolution around the $\Gamma$-point enabled detailed characterization of the isotropic BIC, revealing it to be a quasi-super-BIC.

Moreover, by capturing negative refractive index self-collimation, our results show the potential for sophisticated beam-control strategies in integrated photonic circuits. This observation opens up possibilities for novel photonic devices combining functionalities such as resonator, beam steering, focusing, and reshaping, which would be infeasible using conventional materials.
Remarkably, we  demonstrated a region of frequency where isotropic negative index and negative mass appear around a symmetry-protected BIC in $\Gamma$-point, which can be crucial in the study of the transport of exciton polaritons.

Importantly, our method's inherent scalability and automation suggest its potential as a new standard for photonic characterization.
Eliminating the diffraction and aberration limitations typical of Fourier optics, our reciprocal-space mapping technique offers a robust platform suitable for large-scale fabrication assessment and systematic exploration of photonic structures. 





\section*{Materials and Methods}
\paragraph{Optical Setup}
\hfill \break
Our optical setup comprises: an NKT Photonics SuperK EXTREME EXW-4 high-power supercontinuum white-light laser, an Ocean Optics HR4000 high-resolution spectrometer, and two Standa 8MR190-2 motorized rotation stages.

\paragraph{Sample Fabrication}
\hfill \break
The photonic-crystal sample measures \(1 \times 1\ \mathrm{mm}^2\) and consists of a square lattice of air-filled cylindrical holes. A Si\(_3\)N\(_4\) layer was deposited onto a quartz slide by plasma-enhanced chemical vapour deposition (PECVD), and the hole pattern was defined in ZEP 520 positive electron-beam resist using high-voltage e-beam lithography (Vistec VB300UHR EWF) \cite{Zito2019}. The resist mask was then transferred into the nitride by reactive-ion etching (RIE) in an Oxford Instruments PlasmaLab 80 Plus tool at room temperature, employing CHF\(_3\) and O\(_2\). Lattice parameters were chosen to achieve resonances near 532 nm, with a lattice periodicity $a = 370$ nm, hole radius $r = 0.25a$, and membrane thickness set to $h = 0.22a$, following the design procedure of Ref. \cite{Mocella2015}.

\paragraph{Generation of Experimental Equifrequency Surfaces}
\hfill \break
\textit{Spectral Mapping.}
Experimentally, the sample is first secured at a chosen azimuthal angle $\varphi$ by adjusting one rotation stage, after which the second stage sweeps through the polar angle $\theta$, yielding the band diagram for that fixed $\varphi$. Upon completion of the polar scan, the primary rotation stage advances to the next azimuthal position and the measurement is repeated, producing the corresponding band diagram. By iterating this sequence over a series of $\varphi$ angles, we cover a circular region around the $\Gamma$-point. The acquired data therefore consist of quadruplets $(\lambda,\theta,\varphi,T)$, where $T$ denotes the transmission signal. These measurements are then mapped into reciprocal-space coordinates using the transformation equations below:
\begin{equation*}
    k_x = \frac{2 \pi}{\lambda} \sin{\theta} \cos{\varphi} \quad \text{and} \quad k_y = \frac{2 \pi}{\lambda} \sin{\theta} \sin{\varphi},
\end{equation*}
leading to the generation of EFS, which stacked form 3D DS.

\textit{Noise filtering.} To remove the slowly varying noise from our signal, we first take its 2D Fourier transform.  We then apply a smooth, high-pass mask that gently attenuates the low-frequency content, effectively suppressing broad background variations, while leaving the finer features intact.  After multiplying the mask with the Fourier spectrum, we invert the transform to recover a noise-reduced version of the data.  To further smooth out any residual oscillations or high-frequency noise introduced by the filtering step, we perform a two-pass Savitzky-Golay smoothing.  

\textit{Post-Processing and Visualization.}
The 3D DS were rendered and analyzed using Avizo (Thermo Fisher Scientific), a commercial software package for scientific visualization. Avizo’s volume rendering and EFS extraction tools were employed to highlight topological features such as Dirac points, band degeneracies, and curvature gradients.

\paragraph{Numerical Modeling of Photonic Crystal Dispersion Surfaces}
\hfill \break
\textit{Computational Framework.}
The DS of the asymmetric PhC slab were computed using a custom-developed, fully parallelized Python-based implementation of the rigorous coupled-wave analysis (RCWA) method \cite{Moharam:86}. This in-house code leverages the RCWA algorithm to solve Maxwell’s equations for periodic structures by expanding electromagnetic fields and material permittivity into Fourier harmonics, with optimizations for high-performance computing (HPC)\cite{Xu_SPIE:2023}.

The simulations employed 81 Fourier  harmonics to represent the periodic dielectric profile, ensuring convergence of the electromagnetic solutions. The custom Python code was parallelized using a distributed-memory architecture, enabling efficient domain decomposition across wavevector ($\textbf{k}$) and frequency ($\omega$) grids.

\textit{Dispersion Surface Calculation.}
The 3D DS were generated by stacking iso-frequency contours across the frequency axis. For each $(k_x,k_y)$ point, the eigenfrequencies of the photonic crystal modes were iteratively computed. The wavevector grid spanned the irreducible Brillouin zone, with $k_z$ derived implicitly from the dispersion relation. 

\textit{Numerical Parameters.}
The wavevector components $(k_x,k_y)$ were discretized within the first Brillouin zone with the following resolutions to construct the 3D equifrequency surfaces.
Normalized wavevector resolution: $\Delta k_{N} = 5\times 10^{-4}$, (uniform sampling in $k_x$ and $k_y$), where $k_{N} = k (a / 2\pi)$.
Normalized frequency resolution: $\Delta \omega_{N} = 6\times 10^{-4}$ (step size in angular frequency), where $\omega_{N} = a / \lambda$.

\bibliography{EquiFreq}

\bibliographystyle{Science}

\end{document}